\documentclass[letterpaper,12pt]{article}   
\usepackage{osajnl2} 

\usepackage{setspace}

\usepackage{amsmath,amssymb,amsfonts}
\usepackage{graphics,graphicx}
\usepackage{subfigure}

\newcommand{\order}{\mathcal{O}}

\newcommand{\bD}{\vec {\bf D}}
\newcommand{\bE}{\vec {\bf E}}
\newcommand{\bM}{{\bf M}}
\newcommand{\bL}{{\bf L}}

\newcommand{\bI}{{\bf I}}

\newcommand{\bJ}{\vec {\bf J}}
\newcommand{\br}{\vec {\bf r}}
\newcommand{\bk}{\vec {\bf k}}
\newcommand{\bt}{\vec {\bf t}}
\newcommand{\ba}{\vec {\bf a}}
\newcommand{\bb}{\vec {\bf b}}
\newcommand{\bbf}{\vec {\bf f}}
\newcommand{\dG}{\overline {\overline {\bf G}}_{per}}

\begin{document}

\title{The Rapid Analysis of Scattering from Periodic Dielectric Structures Using Accelerated Cartesian Expansions (ACE)}

\author{Andrew D. Baczewski,$^{1,2*}$ Nicholas C. Miller,$^1$ and Balasubramaniam Shanker$^{1,2}$}
\address{$^1$Department of Electrical and Computer Engineering, Michigan State University, \\ East Lansing, MI, 48825, USA}
\address{$^2$Department of Physics and Astronomy, Michigan State University, \\ East Lansing, MI, 48825, USA}
\address{$^*$Corresponding author: baczewsk@msu.edu}

\begin{abstract*} 

The analysis of fields in periodic dielectric structures arise in numerous applications of recent interest, ranging from photonic bandgap (PBG) structures and plasmonically active nanostructures to metamaterials.  To achieve an accurate representation of the fields in these structures using numerical methods, dense spatial discretization is required.  This, in turn, affects the cost of analysis, particularly for integral equation based methods, for which traditional iterative methods require $\order(N^2)$ operations, $N$ being the number of spatial degrees of freedom.  In this paper, we introduce a method for the rapid solution of volumetric electric field integral equations used in the analysis of doubly periodic dielectric structures.  The crux of our method is the ACE algorithm, which is used to evaluate the requisite potentials in $\order(N)$ cost.  Results are provided that corroborate our claims of acceleration without compromising accuracy, as well as the application of our method to a number of compelling photonics applications. {\it Manuscript submitted to JOSA A.}

\end{abstract*}

\maketitle 

\section{Introduction}
The scattering of light from subwavelength dielectric arrays is fundamental to a number of contemporary problems in optics.  In photonic bandgap (PBG) structures, gaps in the photonic density of states arise due to a modification of the photonic dispersion in the vicinity of Bragg planes brought about by some underlying periodicity \cite{john87}. These structures have been explored for a wide range of applications, from waveguiding structures \cite{lin98} to experimental tests of cavity quantum electrodynamics \cite{yoshie04}.  Similarly, plasmonic structures exploit the unusual dielectric behavior of certain metals (namely, the nobles) at optical frequencies to focus electromagnetic energy at subwavelength scales.  This focusing is typically the result of the excitation of surface plasmon-polaritons (SPP) modes, with dispersion relations that lie below the free space light-line \cite{otto68}.  One means of coupling light into these modes is the periodic modulation of the surface morphology \cite{barnes03}.  This physical phenomenon has been exploited for a number of applications, ranging from extraordinary transmission \cite{ebbesen98} to briding the gap between and electronic and optical circuitry \cite{ozbay06}.  Finally, metamaterial structures have been studied extensively in the hopes of finding media that can be homogenized, albeit at specific wavelengths and angles of incidence, in such a way that they effect the behaviors of a structure with a negative, near zero, or otherwise anomalous index of refraction \cite{shalaev07}.  While the noble metals are also often used in optical metamaterials, the associated losses can be too severe, leading to the development of active metamaterials, wherein dye molecules are used as gain-inducing dopants \cite{xiao10}.

In all of the aforementioned applications, intuition and supplemental simulation have been successfully applied to the design and analysis of countless structures.  However, this approach is bound to fail as designs become increasingly complex.  It is the goal of computational electromagnetics and optics to supplant much of the trial and error associated with the design of such complex structures with rigorous {\it in silico} modeling.  Typical systems that have been previously analyzed using full-wave methods are relatively small in terms of the number of spatial degrees of freedom, ranging from 100s-1,000s.  However, a recent proliferation of research in PBG, SPP, and metamaterial systems with fine subwavelength features, and/or highly lossy/dispersive materials lead to periodic structures wherein the unit cell requires 10,000+ degrees of freedom to reliably represent the underlying physics.

Integral equation (IE) methods for periodic scattering problems present a number of salient advantages over full-wave differential methods.  IE methods give rise to solution fields that exactly satisfy proper boundary conditions through the use of an appropriate Green's function, are free of numerical dispersion, and only require the discretization of scattering bodies.  This is not to say that these methods are without their disadvantages, in particular, these methods give rise to dense matrices that require $\order(N^2)$ resources in terms of number of operations and storage to achieve an iterative solution.  Further, the evaluation of matrix elements by numerical quadrature is somewhat onerous, as the periodic Green's function must be evaluated at each pair of source and testing integration nodes.  To this end, a considerable body of research has developed with the goal of mitigating these costs through the design and implementation of fast algorithms.

Some popular fast methods include the Fast Multipole Method (FMM) \cite{song95}, the Adaptive Integral Method (AIM) \cite{bleszynski96}, and numerous variations that address adaptivity, or low/high frequency bottlenecks.  These methods have become very popular for the analysis of non-periodic scattering problems.  While the framework for adapting FMM to periodic problems in statics has been around since \cite{greengard87}, it was not until the mid-90's that FMM was adapted to periodic wave propagation \cite{rokhlin94}.  Even so, it was used to reduce matrix fill time, rather than accelerating the iterative solution process- this was not realized until the work of Otani and Nishimura in 2008 \cite{otani08}.  Aside from tree-based methods, AIM-based methods have been very naturally employed in the analysis of periodic problems using both conventional IE methods \cite{bleszynski10}, as well as FE-BI \cite{eibert98}.  Further, a number of interpolatory methods have been applied to these problems \cite{shaojing10}, including highly efficient GPU implementations \cite{shaojing10b}.  In this work, we expand upon this growing body of research in fast algorithms for periodic systems, and provide details of the extension of the method of Accelerated Cartesian Expansions (ACE) a hierarchical, tree-based method similar to the FMM, to doubly periodic dielectric arrays.

The ACE algorithm has previously been applied to the efficient computation of potentials of the form $R^{-\nu}$ \cite{shanker07}, Lienard-Wiechert potentials \cite{vikram07}, diffusion, lossy wave, and Klein-Gordon potentials \cite{vikram10}, and periodic Helmholtz, Yukawa, and Coulomb potentials \cite{baczewski11}.  Like FMM, it is based upon a hierarchical decomposition of the computational domain mapped onto an octree data structure, wherein a distinction between near and farfield source-observer aggregates is made, and an addition theorem is used to effect the interaction of all bodies with linear scaling.  Whereas this addition theorem is based upon spherical harmonics in the FMM, ACE utilizes Cartesian harmonics and takes the form of a generalized Taylor expansion.  In doing so, the salient features that develop are: 
\begin{itemize}
  \item {\it Totally linear scaling:} in terms of both computational cost and storage.
  \item {\it Nearly kernel independent framework:} only multipole-to-local operators depend upon the explicit form of the Green's function.
  \item {\it Exact up/down tree traversal:} error is rigorously independent of tree height. 
  \item {\it Amenability to non-uniform discretization:} multiscale structures can be handled very naturally.
  \item {\it Excellent low frequency accuracy:} conventional FMM for Helmholtz problems must be augmented for electrically dense problems \cite{greengard98,jiang05}.  ACE has been shown to be very much complimentary to FMM in this regard \cite{vikram09}.
\end{itemize}   
It is this latter feature that we are particularly interested in exploiting in adapting ACE to periodic structures.  In our primary applications of interest, the unit cells are at most $1-2\lambda$, as the technologically compelling physics arises due to energy coupled into low order Bloch-Floquet modes.  In many applications, the unit cell may be as small as $\lambda/4-\lambda/10$.  

The principal contributions of this work are two-fold, (i) an extension of the ACE algorithm that enables the efficient analysis of electrically dense periodic structures, and (ii) applications of this technology to a set of challenging problems.  In presenting this work we will use the following layout.  In Section 2, we provide a formal mathematical statement of the problem.  In Section 3, we give details of the ACE algorithm, including the details necessary for its extension to volumetric periodic problems.  Section 4 presents results that affirm our claims of linear scaling, convergence of our method to arbitrary accuracy, and utility in accelerating the analysis of a number of exemplary structures.  Finally, Section 5 summarizes the contributions of this paper and gives a brief outline of work to come.

\section{Problem Statement}
Consider a doubly periodic array of dielectric scatterers, $\Omega_D$, embedded in $\mathbb{R}^3$, characterized by the dielectric function $\varepsilon(\br,\omega)$.  The array is excited by an incident planewave, $\bE_{inc}(\br,t)= \bE_0 e^{i(\omega t-\bk_{inc}\cdot\br)}$, that gives rise to a polarization of the dielectric.  In turn, this polarization field radiates a scattered field, $\bE_{scat}(\br,t)$, such that the total field is given by $\bE(\br,t) = \bE_{inc}(\br,t) + \bE_{scat}(\br,t)$.   Considering only linear media, all fields will be time-harmonic, and we henceforth suppress an implicit factor of $e^{i\omega t}$ and all time-dependence.  

\begin{figure}[h]
 \centering
 \includegraphics[scale=0.5]{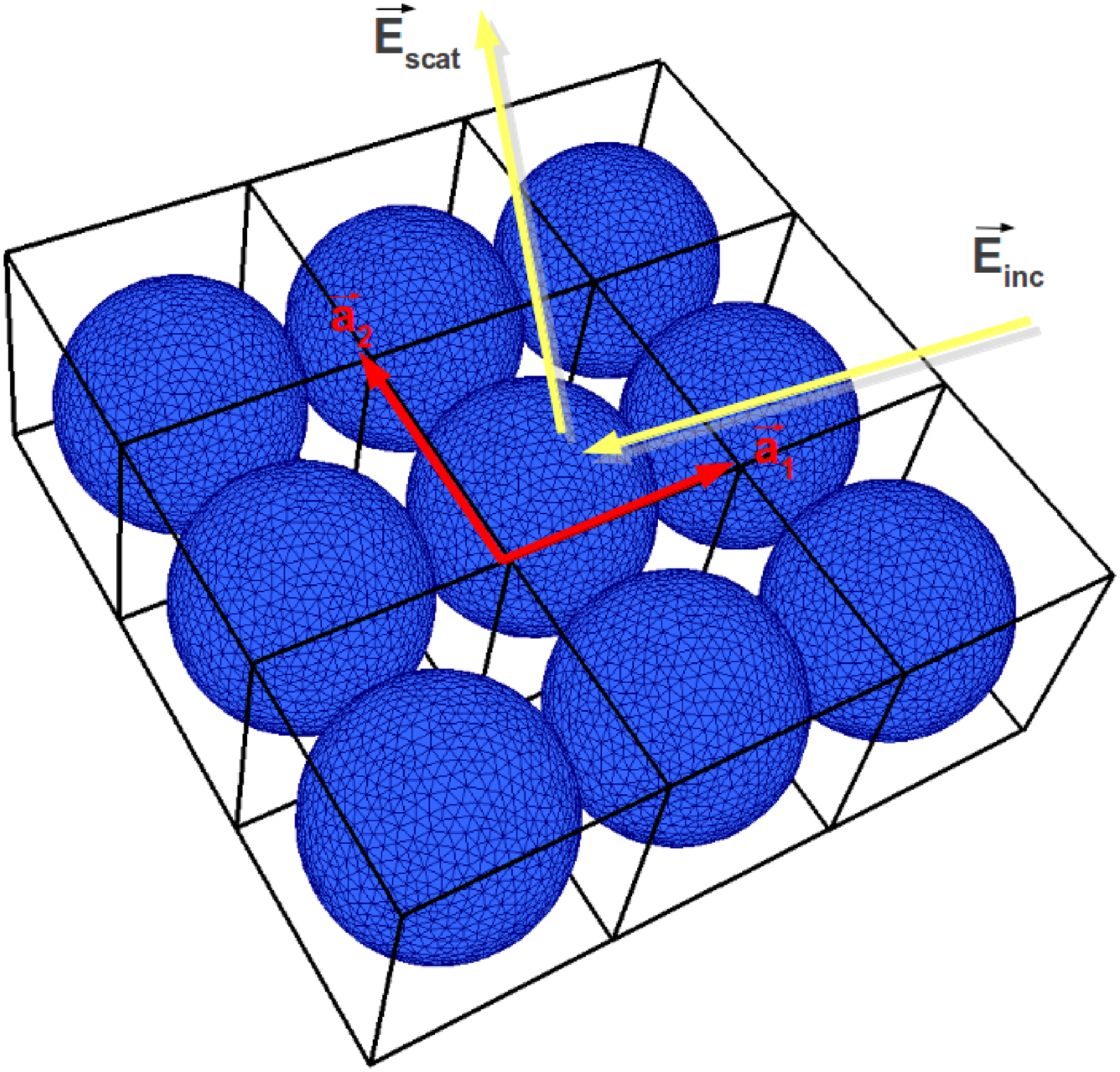}
 \caption{Illustration of the periodic scattering problem described in Section 2.} \label{problem_fig}
\end{figure}

The array upon which the geometry is arranged is characterized by the following 2-lattice, $\mathcal{L}_2$:
\begin{equation}
 \mathcal{L}_2 = \lbrace \bt_{m,n} = m\ba_1 + n\ba_2~|~m,n \in \mathbb{Z}\rbrace
\end{equation}
Here, $\ba_1$ and $\ba_2$ are the lattice vectors, describing the periodicity of our array.  Throughout, we will assume a square lattice, i.e., $|\ba_1|=|\ba_2|$ and $\ba_1\cdot\ba_2=0$, though we note that our method can be extended to more irregular lattices (i.e., rectangular or skewed) with minimal modification.  For completeness, we define the reciprocal lattice associated with $\mathcal{L}_2$ as:
\begin{equation}
 \mathcal{L}^*_2 = \lbrace \bk_{m,n} = m\bb_1 + n\bb_2~|~m,n \in \mathbb{Z}\rbrace
\end{equation}
Where $\ba_i \cdot \bb_j = 2\pi\delta_{ij}$, $\delta_{ij}$ being the Kronecker Delta.  A simple illustration of our problem is provided in Fig. \ref{problem_fig}.

We seek to resolve the unknown, $\bE_{scat}(\br)$, from which we can also compute quantities such as the scattering parameters, e.g., reflection, transmission, and absorption spectra, associated with $\Omega_D$.  In doing so, we utilize the volumetric equivalence principle to replace $\Omega_D$ with equivalent sources, $\bJ_V(\br)$, radiating into a homogeneous space \cite{peterson98}.  We relate $\bJ_V(\br)$ to the electric displacement, $\bD(\br)=\varepsilon(\br,\omega)\bE(\br)$, via the following relationship:
\begin{equation}
 \bJ_V(\br) = j\omega \kappa(\br) \bD(\br),~~~\text{where}~\kappa(\br) = \frac{\varepsilon(\br,\omega)-\varepsilon_0}{\varepsilon(\br,w)}
\end{equation}
With this substitution, any discontinuities in the normal component of $\bJ_V(\br)$ across material interfaces is due to $\kappa(\br)$ rather than $\bD(\br)$, which facilitates the definition of local vector basis functions \cite{schaubert84}.  

By relating $\bE_{scat}(\br)$ to $\bJ_V(\br)$, and enforcing the identity of the total field, we arrive at the following Volume Integral Equation (VIE):
\begin{subequations}
\label{vie}
\begin{align}
  &\bE_{inc}(\br) = \bE(\br) - \bE_{scat}(\br),~~~\forall \br \in \Omega_D \\
  &\bE_{inc}(\br) = \bD(\br)/\varepsilon(\br,\omega) - i\omega\mu_0 \int \limits_{\Omega_D} d\br' g(\br,\br') \kappa(\br')\bD(\br') - \frac{i}{\omega\varepsilon_0}\nabla \int \limits_{\Omega_D} d\br' g(\br,\br') \nabla' \cdot \left(\kappa(\br') \bD(\br')\right) \label{vie_b}
\end{align}
\end{subequations}
Here, $g(\br,\br')$ is the free space Green's function for the Helmholtz equation in 3D.  By taking advantage of periodicity, we restrict our consideration to a single unit cell, $\Omega_{0} \subset \mathbb{R}^3$ , and replace $g(\br,\br')$ with the quasiperiodic Green's function for the Helmholtz equation in 3D, $g_{per}(\br,\br')$. Using $g_{per}(\br,\br')$ a radiation boundary condition is enforced receding away from the array, and Bloch-Floquet boundary conditions are enforced in the plane of $\mathcal{L}_2$, giving the following relationship between fields/currents in different unit cells:
\begin{equation}
 \bE(\br + \bt_{m,n}) = e^{i\bk_{inc}\cdot \bt_{m,n}} \bE(\br)
\end{equation}
Equation (\ref{vie}) can then be reformulated as:
\begin{equation}
 \label{vefie}
 \bE_{inc}(\br) = \bD(\br,\omega)/\varepsilon(\br,\omega) - \int \limits_{\Omega_D^*} d\br' \dG(\br,\br') \cdot \left[ \kappa(\br') \bD(\br') \right],~~~\Omega_D^* = supp(\bJ_V) \cap \Omega_0
\end{equation}
Here, $\dG(\br,\br')$ is the Quasiperiodic Electric Dyadic Green's function, constructed from the scalar Green's function, $g_{per}(\br,\br')$, per the formalism in \cite{harrington01}.  

To render a finite system of equations, $\Omega_D^*$ is represented as a tetrahedral tesselation with $N$ faces, each of which is assigned a linear vector/Schaubert-Wilton-Glisson (SWG) basis function \cite{schaubert84}.  Galerkin testing is subsequently applied \cite{peterson98}, yielding a linear system of $N$ equations in $N$ unknowns of the form:
\begin{subequations}
 \label{mom_eqn}
\begin{align}
 &Z_{\mu \nu} I_{\nu} = V_{\mu}  \label{ohms_law} \\
 &I_\nu = c_\nu,~~~V_\mu = \langle \bbf_\mu(\br), \bE_{inc}(\br)\rangle \\
 &Z_{\mu\nu} = \langle \bbf_\mu (\br) , \bbf_\nu(\br)/\varepsilon(\br,\omega)\rangle - \langle \bbf_\mu(\br), \int d\br' \dG(\br,\br') \cdot \left[\kappa(\br') \bbf_\nu(\br') \right] \rangle \\
 &\bD(\br) = \sum \limits_{i=1}^{N} c_i \bbf_i(\br),~~~\forall \br \in \Omega_D^*
\end{align}
\end{subequations}   
The expansion coefficients for $\bD(\br)$, $c_i$, can be resolved using an iterative method \cite{saad03} in $\order(N^2)$ operations.  These methods require the generation of a minimal sequence of vectors in the range of $Z_{\mu \nu}$ from which an approximate solution to Eqn. (\ref{ohms_law}) can be constructed such that $||Z_{\mu \nu}I_\nu - V_\mu||_2 \leq \epsilon$, where $\epsilon$ is some designated tolerance for error.  This $\order(N^2)$ cost can be reduced via the application of fast solvers, wherein the dominant cost of the iterative solution process is reduced to $\order(N \log^\alpha_2(N))$ for $\alpha \in \left[0, 1\right]$.  In the case of the ACE algorithm, $\alpha=0$, and $\order(N)$ scaling has been demonstrated in numerous applications.  In the Section 3, we give details of how this is achieved for the class of volumetric, doubly periodic vector Helmholtz problems described above.

\section{ACE Algorithm}

The crux of the ACE algorithm is the Generalized Taylor Expansion (GTE) expressed in the framework of Cartesian tensors:
\begin{equation}
 f(\br-\br') = \sum \limits_{n=0}^{\infty} \frac{(-1)^n}{n!} \br'^{(n)} \cdot n \cdot \nabla^{(n)} f(\br)
\end{equation}
Here, $f(\br)$ is some smooth function, $\cdot n \cdot$ indicates an $n$-fold tensor contraction, and the superscript $(n)$ indicates a tensor of rank $n$.  The premise of this equation, and of the ACE method, is to utilize the GTE in constructing an addition theorem for the periodic Green's function, wherein the source ($\br'$) and observer ($\br$) domains are separated.  As in other analytical acceleration methods, it is this separation that enables the contruction of a hierarchical decomposition of the potential integrals (rendered discretely as matrix-vector products) that are repeatedly evaluated in the iterative solution of Eqn. (\ref{mom_eqn}).  To use this principle in achieving a linear scaling algorithm, we need to first designate when it is that we will utilize this separation to effect a portion of a matrix-vector product (i.e., nearfield vs. farfield metric), and then construct a mathematical framework that describes the process (i.e., tree traversal) in which it is applied.  To this end, we describe:
\begin{itemize}
 \item Construction of a hierarchical decomposition of $\Omega_D^*$ and its utility in identifying nearfield vs. farfield basis function pairs.
 \item Application of the tree traversal process to effect a matrix-vector product.
\end{itemize}

\subsection{Tree Construction}
\begin{figure}[h]
 \centering
 \includegraphics[scale=0.5]{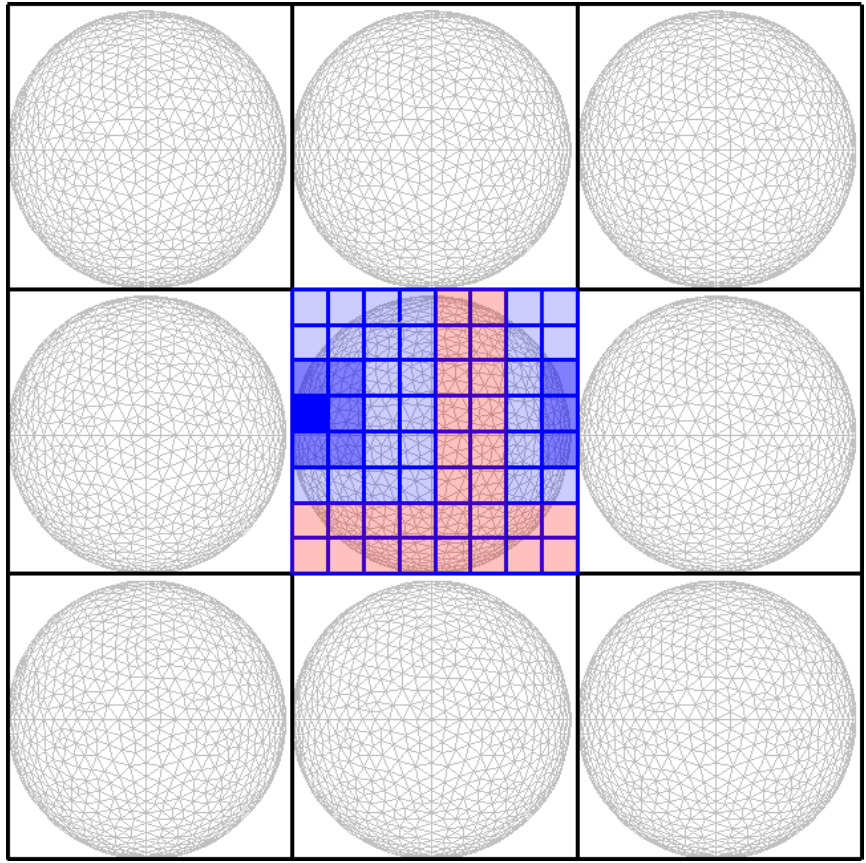}
 \caption{Top-down view of the geometry illustrated in Fig. \ref{problem_fig} with a 4 level octree structure superimposed.  Interaction lists are indicated for the dark blue box.  Boxes highlighted in blue are in the nearfield, whereas light blue boxes are in the farfield.  The interaction between sources in the dark blue box and boxes highlighted in red are effected at a higher level.\label{int_list}}
\end{figure}

In the ACE algorithm, the matrix-vector product in Eqn. (\ref{ohms_law}) is approximated as:
\begin{equation}
 Z_{\mu \nu} I_\nu \approx Z^{near}_{\mu\nu} I_\nu  + \mathcal{T}^{ACE}(I_\mu) 
\end{equation}
Here, $Z^{near}_{\mu\nu}$ is a sparse matrix with $\order(N)$ entries which describe only interactions between pairs of basis functions that are in some metric near.  The remaining term, $\mathcal{T}^{ACE}$ is some composition of operators that effects the remaining farfield contribution to the total in $\order(N)$ cost.  The distinction between nearfield and farfield is made apparent by mapping the basis functions subordinate to $\Omega_D^*$ onto a hierarchical decomposition, i.e., a tree.  This is accomplished by embedding the unit cell inside of a cubic box and recursively subdividing it into increasingly smaller cubic boxes until some desired density of basis functions per box, $\sigma$, is achieved.  This entire hierarchy of boxes is stored and represented in terms of a regular octree data structure.  Boxes are referred to in terms of their genealogy, e.g., the box subordinate to a box at the level directly above it is the child to the superordinate box's parent.  This genealogy is used in constructing the nearfield/farfield dictum as follows - the basis functions subordinate to two boxes constitute a farfield pair if:
\begin{itemize}
 \item The separation between the original boxes and their nearest periodic images is, in all cases, at least a box length.
 \item Among the parents of both the original boxes and their images, at least one pair are in each other's nearfield.
\end{itemize}
The tree structure, and an exemplary application of this dictum are illustrated in Fig. \ref{int_list}. 

It is important to note that the designation of nearfield and farfield does not carry the same physical significance as it typically does, but is only used in analogy with conventional free space tree-based methods.  In practice, we implement this augmented rule by simply adding 2 auxiliary levels to our tree to account for the 8 nearest images of the unit cell, which are addressed but not explicitly `filled'.  As we utilize Morton ordering, the boxes lying in the original unit cell lie inside of a contiguous address space and we can easily distinguish between image boxes and the original boxes, but still utilize the same list building routines as in a free space code.  This augmentation has a negligible effect on the computational cost, and can be considered a step in pre-processing as the image boxes are not explicitly traversed.

\subsection{Tree Traversal}
Next, we discuss the application of $\mathcal{T}^{ACE}$.  As the ACE algorithm is formulated in the language of Cartesian tensors, it is necessary to first provide notational details.  $\bM_\mu^{(n)}$ and $\bL_\mu^{(n)}$ are deemed the multipole and local expansions.  These quantities are $3$-vectors, indexed by $\mu$, each component of which is a totally symmetric Cartesian tensors of rank $n$.  A rank $n$ Cartesian tensor on $\mathbb{R}^3$ is a set of $3^n$ quantities indexed by $x$, $y$, and $z$ in each rank.  Such a tensor is totally symmetric if these quantities are left invariant under permutation of the indices across rank, and will thus consist of $n(n+1)/2$ unique quantities.  We denote an $n$-fold contraction between such tensors as $\cdot n \cdot$, and any vector quantity such as $\nabla$ or $\br$ taken as an $n$-fold tensor product with itself is indicated as $\nabla^{(n)}$ or $\br^{(n)},$ respectively.  Further details concerning tensor notation, as well as proofs of the identities that will follow can be found in \cite{shanker07}.  

For simplicity, we restrict ourselves to a 3-level tree, in which there will be no up/down tree traversal, noting that the details of a multilevel implementation can be found in \cite{shanker07, vikram09}.  We begin by considering two disjoint subdomains of $\Omega_0$, $\Omega_s$ and $\Omega_o$ with centroids $\br^c_s$ and $\br^c_o$, deemed the source and observer domains, respectively.  These subdomains are spatially separated in such a way that they separately fall inside leaf boxes that satisfy the farfield criterion described in Section $3.A$.  The first step in tree traversal is the construction of ACE multipole expansion about $\br^c_s$, commonly referred to in the literature as the `charge-to-multipole' or C2M step.  For an ACE expansion truncated beyond order $P$, this amounts to calculating the $\order(P^3)$ unique entries of the following set of totally symmetric Cartesian tensors:
\begin{equation}
 \bM_\mu^{(n)} = \sum \limits_{i=1}^{N_s} \frac{(-1)^n}{n!} w_{\mu,i} (\br_i - \br_s^c)^{(n)} ,~  0 \le n \leq P
\end{equation}
Here, $N_s$ is the number of quadrature points used in discretizing the contribution of the source integrals in Eqn. (\ref{mom_eqn}) due to basis functions in $\Omega_s$, and $w_{\mu,i}$ is the weight associated with the $\mu$th vector component of the $i$th quadrature point.  

The coefficients contained in the $\bM_\mu$ tensors, combined with a knowledge of the Taylor coefficients of the Green's function allow us to calculate the fields at any point in $\br^c_o$, from a local expansion about $\br^c_o$, defined as:
\begin{equation}
 \bL_{\mu}^{(n)} = \sum \limits_{m=n}^{P} \frac{1}{n!} \bM_{\mu}^{(m-n)} \cdot (m-n) \cdot \nabla^{(m)} g_{per}(|\br^c_o-\br^c_s|),~ 0 \le n \leq P
\end{equation}
Here, the afforementioned Taylor coefficients are contained in the set of totally symmetric tensors, $\nabla^{(m)} g_{per}(|\br_o^c - \br_s^c|), 0\leq m \leq P$, referred to as the multipole-to-local (M2L) translation operator.  As the quasi-periodic Green's function is an infinite sum, an efficient means of calculating the elements of the translation operator is necessary.  To this end, the Ewald representation of the quasi-periodic scalar Helmholtz Green's function is used to generate these coefficients, a more detailed discussion of which can be found in \cite{baczewski11}.  We will make clear how the dyadic character of the Green's function used in the VIE can be accounted for using the scalar Green's function in what follows.

The final step in computing the fields in $\Omega_o$ due to sources in $\Omega_s$, $\bE_\mu^{so}(\br)$, is termed the local-to-observer (L2O) step.  This essentially amounts to the evaluation of the last two terms on the right hand side of Eqn. (\ref{vie_b}).  These fields are evaluated using the following relationship:
\begin{equation}
 \label{dyad_eqn}
 \bE^{os}_\mu(\br) = \left(\bI^{(2)} + \frac{\nabla^{(2)}}{k^2} \right) \cdot 1 \cdot  \sum \limits_{n=0}^{P} \bL_\nu^{(n)} \cdot n \cdot (\br - \br_o^c)^{(n)},~\br \in \Omega_o
\end{equation} 
Here, $\bI^{(2)}$ is the rank $2$ identity tensor.  As $\bL_\mu^{(n)}$ is a constant and the source and observer domains are disjoint, $\nabla^{(2)}$ can be applied directly to the $(\br-\br_o^c)^{(n)}$ tensors.  From these fields, the testing integrals in Eqn. (\ref{mom_eqn}) can be evaluated, completing the farfield contribution to the matrix-vector multiplication.  It is worth noting, that we can alternatively evaluate the tested field by transferring the $\nabla^{(2)}$ onto the source and testing basis functions prior to tree traversal.  In this case, it is necessary to traverse an additional tree to account for the static contribution to the field.  In practice, the difference between these two approaches comes down to a trade off between speed and accuracy.  Transferring the derivatives typically results in a slightly more accurate evaluation of the field, whereas evaluating the field using the dyadic relation in Eqn. (\ref{dyad_eqn}) is much more efficient.  Consequently, the results presented in this paper have been obtained using the latter strategy.

\subsection{Computational Cost}
The cost structure of this scheme is identical to the one presented in \cite{baczewski11}, with the exception of some minor details.  In \cite{baczewski11} it is demonstrated that this periodized version of ACE provides $\order(N)$ evaluation of scalar potentials in terms of both FLOPs and memory.  There are two primary differences between the scalar potentials analyzed previously and the vector potentials presented in this paper.  First, a separate tree must be traversed for each vector component of the field, which will increase the number of operations and memory required for a tree traversal by a factor of either $3$ or $4$ depending upon whether or not the static contribution to the potential is evaluated using a separate tree, as previously discussed.  Second, the quantity $N$ is no longer the number of degrees of freedom (i.e, basis functions in this case), but proportional to this quantity.  Instead, it is determined by the source and testing quadrature rules used in discretizing the necessary MoM integrals.  While this doesn't affect the $\order(N)$ scaling of the method, it does affect the optimal number of degrees of freedom per leaf box used in minimizing cost.  In the results presented, the average number of basis functions per leaf box is $\sim 20$.

\section{Results}

Next, we present several results that:
\begin{itemize}
 \item Validate the proposed acceleration scheme and demonstrate convergence.
 \item Demonstrate the applicability of our method to the analysis of a numerical of established results.
 \item Explore the utility of the algorithm in analyzing a number of interesting photonic structures.
\end{itemize}

\begin{figure}[h]
 \centering
 \includegraphics[scale=0.5]{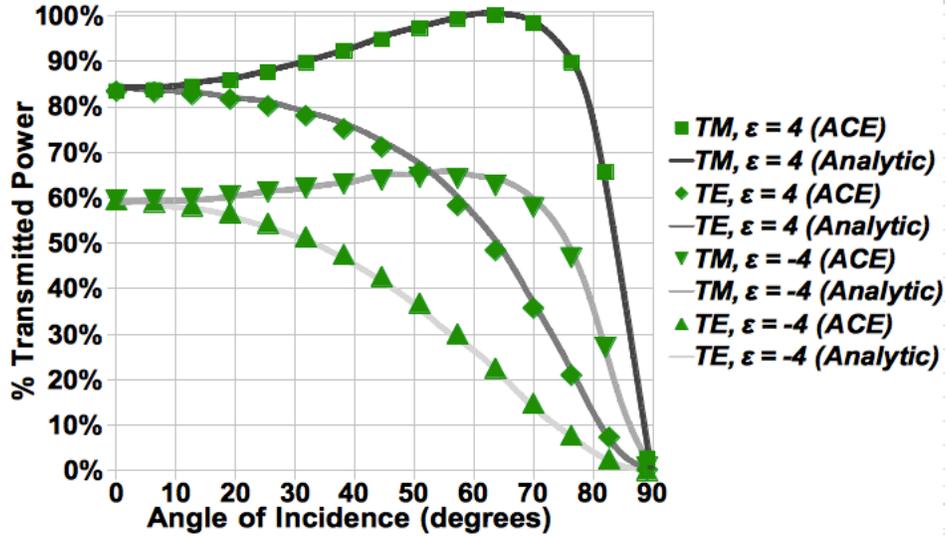}
 \caption{Validation of our ACE accelerated code against an analytic solution for scattering from a homogeneous dielectric slab of width $20$ nm with $\varepsilon_r=\pm 4$. \label{analytic_val}}
\end{figure}

Both our ACE accelerated and reference codes were written in Fortran.  They were compiled using the Intel Fortran Compiler with $-r8 -O3$ flags (double precision and optimization), and run in serial on an Intel Xeon E5620 at 2.4 GHz with 24 GB of RAM running Linux OS at the High Performance Computing Center at Michigan State University.  In generating all results, we utilize the a diagonal-preconditioned TFQMR iterative method with a tolerance of $< 0.1\%$ in the $L_2$ norm.  This choice of norm is implicit throughout the remainder of this work.  All integrals over tetrahedra are evaluated with a 5 point source, 1 point testing rule, whereas all integrals over triangles are evaluated with a 7 point source, 1 point testing rule.  This is of course, excepting singular integrations, which are evaluated using an analytic singularity subtraction in conjunction with quadrature \cite{wilton84}.  All periodic Green's function evaluations, in both Ewald and Floquet representations, are converged to an accuracy of $<0.001\%$.  All infinite series in the periodic ACE translation operators are evaluated to a similar accuracy.  In all cases, we note that we can significantly reduce the runtime of our algorithm by reducing the order of our ACE expansions and/or the convergence criterion for infinite summations.  Unless otherwise indicated, we have utilized $5$ levels and $P=7$ to guarantee an error of $\sim 0.01\%$, in line with the results presented in \cite{baczewski11} and in this work.  In most scattering-based applications, a lower error tolerance/order of ACE expansion can be used with minimal recourse to the computed farfield observables.  Finally, all exciting fields are normally incident on the plane of periodicity, excepting the results in Fig. (\ref{analytic_val}).

The first two results are obtained using a `kernel code' from which error convergence and scaling can be extracted independent of the framework of an integral equation solver.  This `kernel code' evaluates the following convolution, both with and without ACE:
\begin{subequations}
\begin{equation}
 \label{potential}
 \Phi(\br) = \int \limits_{\Omega_0} d\br' g_{per}(\br,\br') f(\br')
\end{equation}
\begin{equation}
 f(\br') = \sum \limits_{i=1}^{N} w_i \delta(\br'-\br_i)
\end{equation}
\end{subequations}
Here, $N$ is the number of co-located point sources/observers, and $w_i$ and $\br_i$ denote the weight and position associated with the $i$th source, respectively.  We are primarily concerned with the farfield contribution to this potential, i.e., for a fixed point observer, $\br_i$, we restrict the integration in Eqn. (\ref{potential}) to include only contributions from sources that are in the farfield of the observer as defined by the ACE algorithm, independent of whether or not ACE is used to evaluate the convolution.  We denote this partial contribution to $\Phi(\br)$ as $\Phi_{far}(\br)$.  

\begin{figure}[h]
\centering
\includegraphics[scale=0.55]{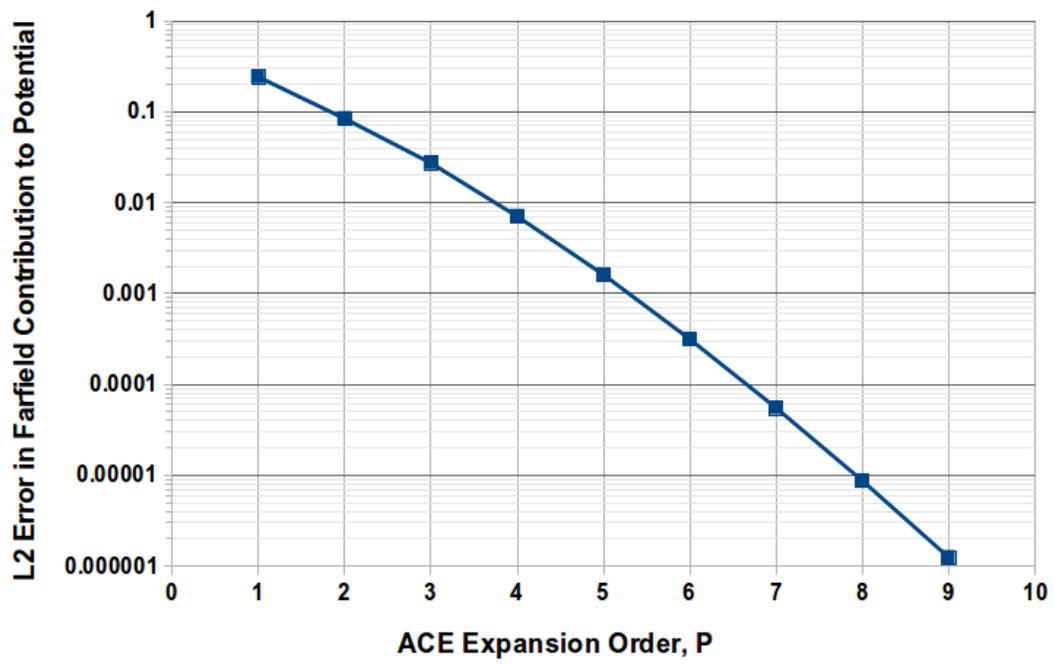}
\caption{Error convergence for $\Phi_{far}(\br)$ evaluated using Eqn. (\ref{error_eqn}).  \label{pconv_fig}}
\end{figure}

In Fig. (\ref{pconv_fig}) error convergence with the order of the ACE expansion is demonstrated using the `kernel code'.  Here, the error was calculated using the following formula for $9$ different values of $P$:
\begin{equation}
 \label{error_eqn}
 \epsilon_{far} = \sqrt{\frac{\sum \limits_{i=1}^{N} ||\Phi^{ACE}_{far}(\br_i) - \Phi^{direct}_{far}(\br_i)||^2}{\sum \limits_{i=1}^{N} ||\Phi^{direct}_{far}(\br_i) ||^2}}
\end{equation}
In the results presented, $N=1,000$ point sources were distributed randomly throughout a unit cell, $\Omega_0 = \left[0, 1\right) \times \left[0, 1\right) \times \left[0, 1 \right)$ (i.e., the unit cube) and the weights, $w_i$, were chosen from a uniform random distribution on $\left[ -1, 1 \right)$.  The phasing in $g_{per}(\br,\br')$ is characterized by an incident field with $\bk_{inc} = 2\pi \ba_1 \times \ba_2$, i.e., it is normally incident on the periodic array and has $\lambda =1$.  The results presented in Fig. (\ref{pconv_fig}) indicate that the ACE algorithm can be tuned to an arbitrary level of accuracy by increasing the order at which the associated expansions are truncated.  We have previously demonstrated the effect of changing incidence angle and wavelength on error for a fixed value of $P$ \cite{baczewski11}, and note here, that it has been found that convergence is largely unaffected by incidence angle, but not wavelength.  Relative to the results presented, as wavelength decreases, convergence in $P$ is slower, whereas it is typically faster as wavelength is increased.  The conditions for this numerical experiment were  chosen to demonstrate convergence in something of a worst case scenario for the applications at hand - as most of the technologically compelling physics for PBGs, plasmonic nanostructures, and metamaterials arise when the unit cell is subwavelength.  It is worth noting that $\epsilon_{far}$ is evaluated relative to $\Phi_{far}(\br)$, rather than $\Phi(\br)$.  As the nearfield contribution to the $\Phi(\br)$ is evaluated exactly, the error with respect to the total potential is typically an order of magnitude smaller.

\begin{figure}[h]
\centering
\subfigure[Scaling of precomputation time with the number of point sources for expansions truncated at different values of $P$]{
\includegraphics[scale=0.5]{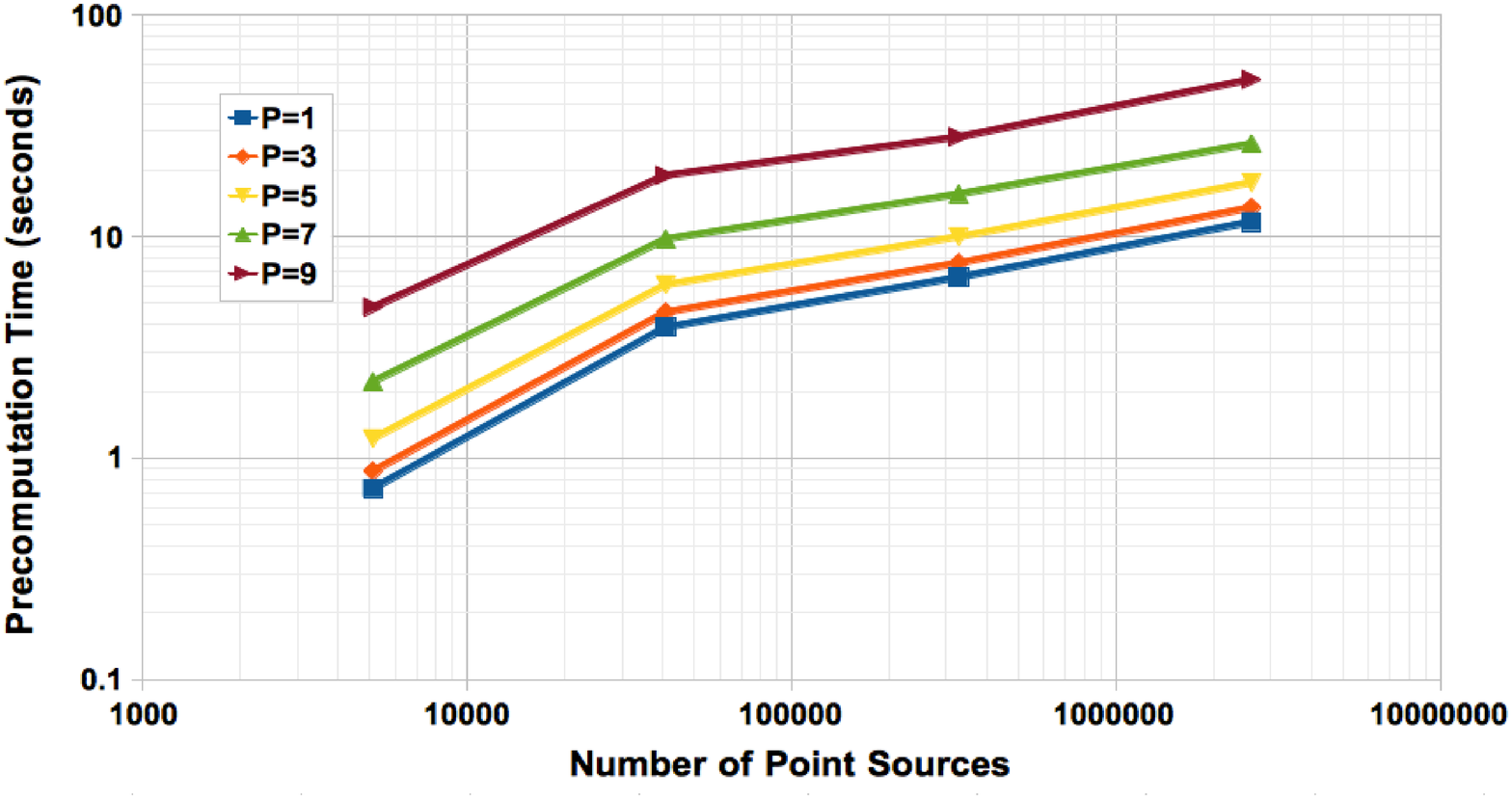}
\label{pre_fig}
}
\subfigure[Scaling of tree traversal time with the number of point sources for expansions truncated at different values of $P$.]{
\includegraphics[scale=0.48]{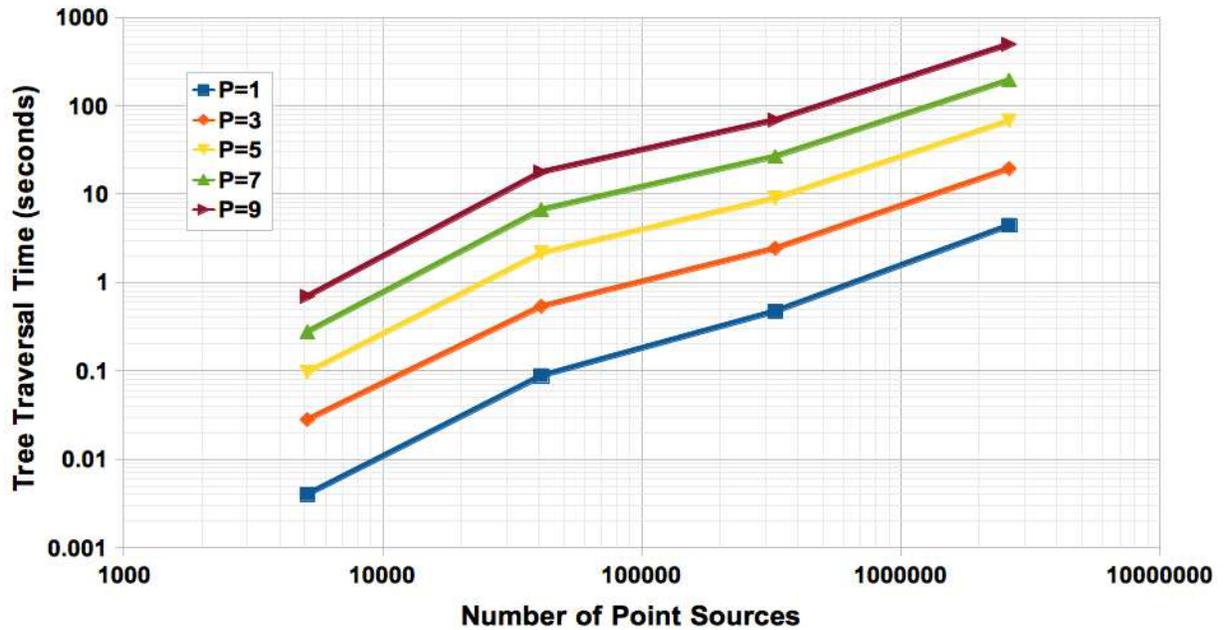}
\label{trav_fig}
}
\caption{Timing results for ACE `kernel code'.\label{timing_fig}}
\end{figure}

The second set of `kernel code' results are given in Fig. \ref{timing_fig}.  Here, scaling of the precomputation and tree traversal times with the number of points sources $N$ are presented.  Timings are presented for 5 different values of $P$, demonstrating that linear scaling can be achieved at arbitrary precision.  In generating the results in this Figure, $N$ sources are distributed over a unit cell with $|\ba_1|=|\ba_2|=1$, with a maximum out-of-plane dimension of $1$, and $k=2\pi$.  The number of levels is increased at each value of $N$, from 3 to 6, maintaining an average density of unknowns per leaf box of $80$.  The precomputation timings in Fig. \ref{pre_fig} include the time required for tree construction as well as the calculation of all unique translation operators that will be later applied during tree traversal.  This is a one time cost, and figures are provided to demonstrate two things: i) precomputation scales weakly with the number of sources/number of levels and ii) precomputation is negligible on the time scale required for solution, as will be evident from subsequent results.  The tree traversal timings in Fig. \ref{trav_fig} include the time required for all 5 steps of the tree traversal project.  These results are intended to demonstrate that linear scaling is achieved for varying levels of accuracy (i.e., different values of $P$).  A linear regression yields a scaling of $N^{1.01}$ for all values of $P$ except for $P=1$, in which the scaling is $N^{1.09}$. 

Our first result illustrating integration with a VIE solver is given in Fig. \ref{analytic_val}.  In it, we demonstrate the accuracy of our method in reproducing power transmission through an infinite dielectric slab of width $20$ nm as the angle of incidence and polarization is varied at a fixed wavelength, $\lambda = 400$ nm.  A similar result is presented in \cite{kobidze05}, wherein a layered-medium formulation is compared against an analytic calculation.  Here, we utilize a periodic volumetric formulation wherein the unit cell has an edge length of $80$ nm.  As is evident, our ACE accelerated code reliably reproduces the analytic solution at both TE and TM polarization, at arbitrarily shallow incidence, for both positive and negative relative permittivities.

\begin{figure}[h]
 \centering
 \includegraphics[scale=0.45]{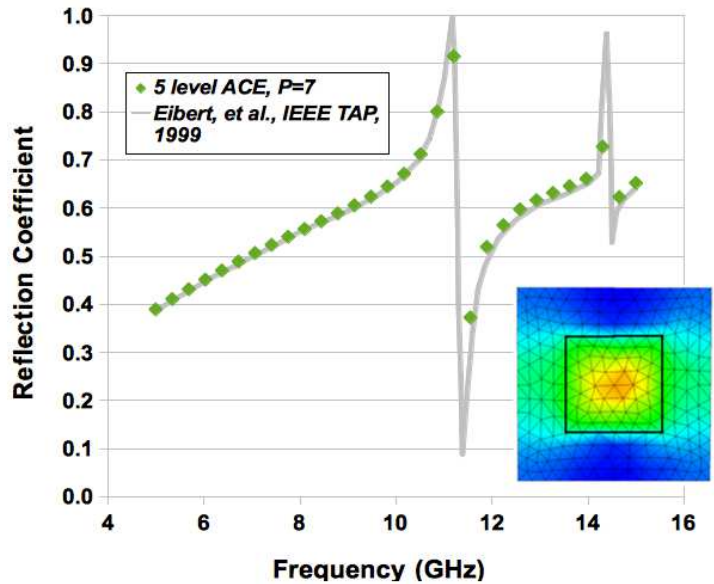}
 \caption{Validation against scattering from an electromagnetic bandgap (EBG) structure solved using FE-BI in \cite{eibert99}.  Discretization has $N=6,030$ unknowns. Average time to solution (without acceleration): $\sim 813$ minutes. Average time to solution (ACE acceleration): $\sim 23$ minutes. Total speedup: $\sim 37 \times$. \label{volakis}}
\end{figure}

\begin{figure}[h]
 \centering
 \includegraphics[scale=0.5]{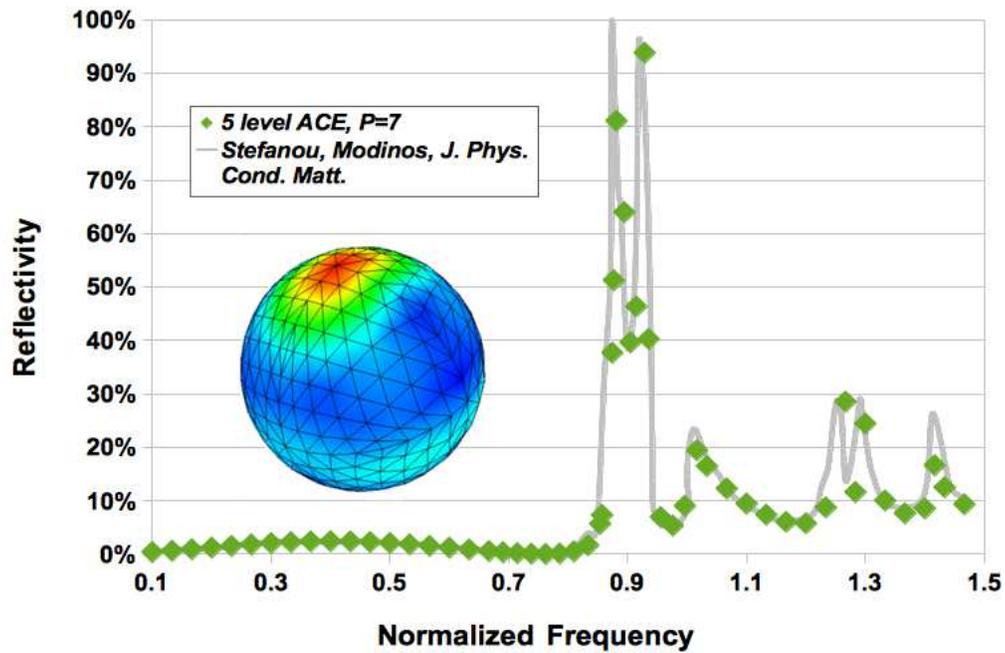}
 \caption{Validation against scattering from an array of polystyrene spheres ($\varepsilon_r=2.56$) solved analytically in \cite{inoue82}.  Discretization has $N=7,328$.  Average time to solution (without acceleration): $\sim 2006$ minutes.  Average time to solution (ACE acceleration): $\sim 43$ minutes.  Total speedup: $\sim 46\times$. \label{stefanou}}
\end{figure}

The results in Fig. \ref{volakis} and Fig. \ref{stefanou} are intended to illustrate i) agreement of our code with established results from the literature and ii) acceleration of our code relative to our own unaccelerated reference code.  The structure in Fig. \ref{volakis} is an electromagnetic bandgap structure with applications in microwave engineering.  The reference data is taken from \cite{eibert99}, wherein a hybrid finite element-boundary integral formulation is utilized.  Our ACE accelerated code is in good agreement with the established result, and we report a $37\times$ speedup relative to our unaccelerated code.  The result in Fig. \ref{stefanou} is taken from \cite{inoue82} wherein an analytic calculation is performed using Mie theory.  We again find good agreement between ACE and the established result and report a $46\times$ speedup relative to our reference code.  This structure has been used elsewhere in the literature as validation of a periodic FMM code, wherein the authors report $4-18$ minutes per solution on an 8 processor platform \cite{otani08}.  As a surface integral equation formulation is utilized, as well as parallelism, it is difficult to draw a direct comparison between our results.  Taking these differences into consideration, it is safe to at least claim that our timings are competitive with the state of the art.

\begin{figure}[h]
 \centering
 \includegraphics[scale=0.2]{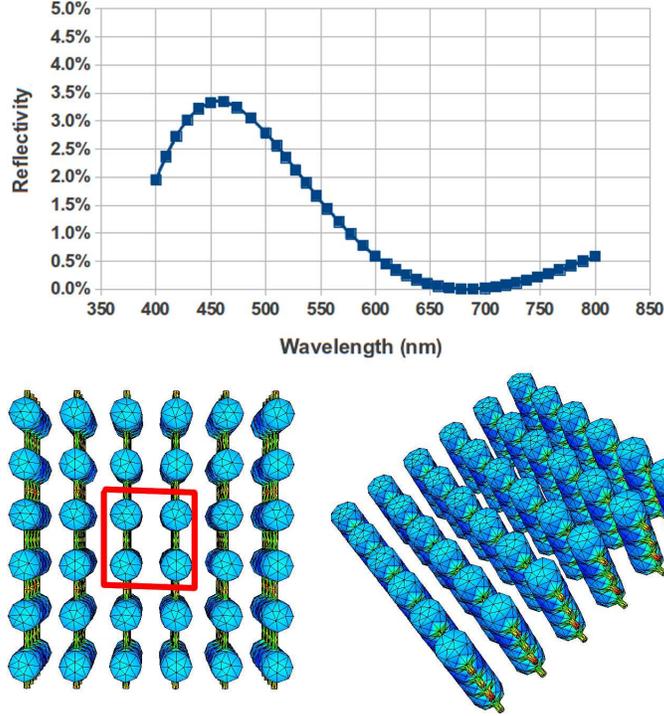}
 \caption{(Top) Calculated reflectivity of a single model butterfly scale. (Bottom) Scale geometry. A single unit cell with $|\ba_1|=|\ba_2|=320$ nm is outlined in red.  The height of the structure out of plane is $280$ nm and the diameter of the larger cylinders is $130$nm, with a center-center spacing of $160$ nm between nearest neighbors. The smaller cylinders of diameter $20$ nm with axes in the plane of periodicty are oriented along the polarization vector of the incident field, with a center-center spacing of $70$ nm out of the plane of periodicity.  The resultant mesh has $N=10,782$.  Average time to solution (without acceleration, extrapolated): $\sim 1430$ minutes.  Average time to solution (ACE acceleration): $\sim 31$ minutes.  Total speedup: $\sim 46 \times$. \label{butterfly_result}}
\end{figure}

Our next result is intended to demonstrate the utility of the ACE algorithm in studying models for novel experimental applications.  It has been understood for many years that a number of the optical effects common in the wings of butterflies, such as their deep coloring or iridescence, arise due to periodic nanostructures that occur naturally in the scales covering these wings, on top of any chemical coloring (i.e. pigment) that may exist \cite{srinivasarao99,vukusic03,parker07}.  The blue coloring of the {\it Morpho} butterfly is partially due to a photonic bandgap (PBG) in the blue portion of the visible spectrum.  This PBG is supported by a periodic ribbing in the material covering the butterfly's wings on the scale of visible wavelengths \cite{srinivasarao99}.  Previously, Huang, et al.\cite{huang06} have performed an extensive analysis of the optical properties of not just naturally occuring wings, but coated and synthetic wings made from Al$_2$O$_3$.  We have constructed a simple model to replicate the blue structural coloring of these types of structures, albeit one that is both thinner and oriented differently than the one in \cite{huang06}.  The reflectance of this structure, as well as its dimensions are given in Fig. \ref{butterfly_result}.  Dielectric constants for Al$_2$O$_3$ were interpolated from \cite{palik91} and the resultant mesh has $N=10,782$ unknowns.  It is worth noting that we have artificially enlarged the unit cell for this problem to ensure that it is cubic.  This both simplifies tree construction and increases the number of degrees of freedom for demonstrative purpose, and is not, in principal, necessary.  In this result, the unaccelerated time to solution was extrapolated based upon the nearfield matrix fill time for ACE alone, neglecting the time for iterative solution, so the speedup factor of $46 \times$ is actually representative of a lower bound on the acceleration.   

Our final result is a demonstration of capability in solving a very densely discretized structure with a highly dispersive dielectric response.  For inspiration, we turn to \cite{xiao09} wherein a yellow-light fishnet metamaterial made of layered Al$_2$O$_3$ and silver on an ITO substrate is simulated and fabricated.  Here, we model a structure with the same periodicity and nanostrip widths but without tapering and an unmodified dielectric function for silver.  While we utilize the well-established Johnson and Christy permittivity for silver \cite{johnson72}, the authors of \cite{xiao09} increase the collision rate to account for surface roughness, grain boundaries, and size effects, thus we do not compare our results directly.  Our resultant mesh has $N=147,374$ unknowns, and was solved using ACE with $6$ levels and $P=5$.  TFQMR was applied with diagonal preconditioning to arrive at an error of $<3\%$ with an average of $188$ iterations per frequency at 20 frequencies.  The average time per iteration was $3.18$ minutes, and the average total time to solution was $896$ minutes.  Extrapolating our nearfield matrix fill time indicates that the matrix fill process alone would require $256,500$ minutes using our reference code with the same parameters, yielding a lower bound on the speedup of $\sim 286 \times$.

\section{Conclusion}
We have demonstrated the extension of the ACE algorithm to the efficient analysis of volumetric integral equation-based scattering problems on periodic domains.  Results were presented that indicate convergence of the kernel to arbitrary precision, as well as linear scaling in the evaluation of matrix-vector products.  Corroborating results affirming the utility of our method in solving problems from the literature were also provided, as well as a number of large problems that demonstrate capability in solving problems posed on densely discretized geometries pertinent to a number of subfields of the optics community.  Work is presently under way in adapting our algorithm to both MPI and CUDA parallelism, surface formulations for dielectrics using the Generalized Method of Moments \cite{nair11}, and time domain problems.

\section*{Acknowledgments}
The authors thank the National Science Foundation Graduate Research Fellowship (ADB) and the Michigan State University Summer Research Expereience for Undergraduates (NCM) programs for supporting their research.  More generally, the authors would like to acknowledge the body of work funded by NSF CCF-0729157 and NSF DMS-0811197 that have also made this work possible, as well as conversations and codes from Dan Dault, He Huang, Naveen Nair, and Melapudi Vikram.  Finally, we thank the High Performance Computing Center (HPCC) at Michigan State University for access to computational resources.

\bibliographystyle{osajnl}

\begin{figure}[h]
 \centering
 \includegraphics[scale=0.4]{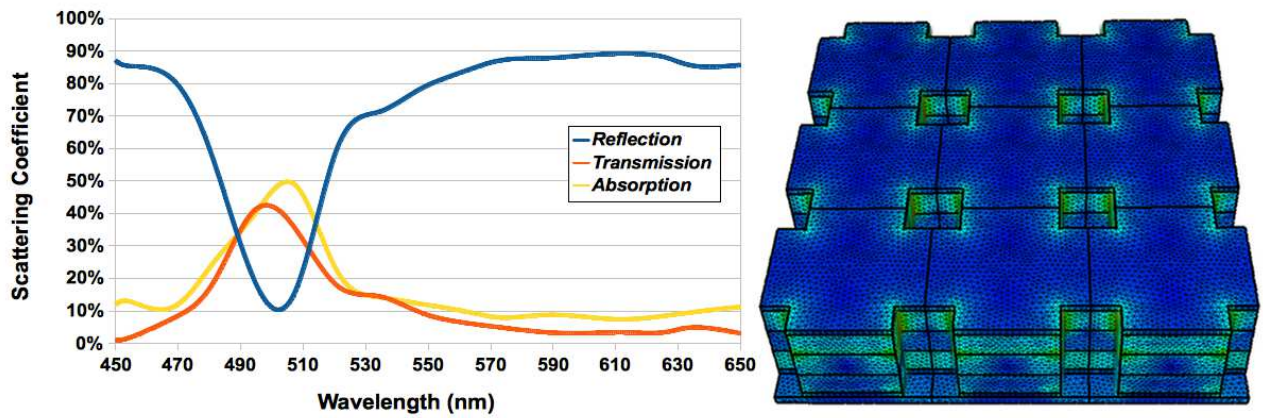}
 \caption{Demonstration of capability in solving a large scattering problem ($N=147,374$) inspired by a metamaterial design presented in \cite{xiao09}.  An average of $188$ iterations per frequency was required, with an average time per iteration of $3.18$ minutes, and an average total solution time of $896$ minutes.}
\end{figure}

\end{document}